\documentclass[bibyear]{aa}
\usepackage{graphicx}
\usepackage{txfonts}

\newcommand{\whya}{\mbox{W~Hya}}
\newcommand{\RSTAR}{\mbox{$R_{\star}$}}
\newcommand{\LSOL}{\mbox{$L_{\sun}$}}
\newcommand{\RSOL}{\mbox{$R_{\sun}$}}
\newcommand{\MSOL}{\mbox{$M_{\sun}$}}
\newcommand{\MSOLPERYR}{\mbox{$M_{\sun}$~yr$^{-1}$}}
\newcommand{\micron}{\mbox{$\mu$m}}
\newcommand{\corundum}{\mbox{Al$_2$O$_3$}}
\newcommand{\forsterite}{\mbox{Mg$_2$SiO$_4$}}
\newcommand{\enstatite}{\mbox{MgSiO$_3$}}
\newcommand{\Ha}{\mbox{H$\alpha$}}
\newcommand{\HOH}{\mbox{H$_2$O}}
\newcommand{\wlc}{\mbox{$\lambda_{\rm c}$}}
\newcommand{\Qplus}{\mbox{$Q_+$}}
\newcommand{\Qminus}{\mbox{$Q_-$}}
\newcommand{\Uplus}{\mbox{$U_+$}}
\newcommand{\Uminus}{\mbox{$U_-$}}
\newcommand{\IQplus}{\mbox{$I_{Q_+}$}}
\newcommand{\IQminus}{\mbox{$I_{Q_-}$}}
\newcommand{\IUplus}{\mbox{$I_{U_+}$}}
\newcommand{\IUminus}{\mbox{$I_{U_-}$}}
\newcommand{\IP}{\mbox{$I_{\rm P}$}}
\newcommand{\PL}{\mbox{$p_{\rm L}$}}
\newcommand{\alfcenA}{\mbox{$\alpha$~Cen~A}}

\begin{document}

\title{
Clumpy dust clouds and extended atmosphere of the AGB star W~Hya 
revealed with VLT/SPHERE-ZIMPOL and VLTI/AMBER
\thanks{
Based on SPHERE and AMBER observations made with the Very Large Telescope 
and Very Large Telescope Interferometer of the European Southern Observatory. 
Program ID: 095.D-0397(D) and 093.D-0468(A). 
}
}

\author{K.~Ohnaka\inst{1} 
\and
G.~Weigelt\inst{2} 
\and
K.-H.~Hofmann\inst{2} 
}

\offprints{K.~Ohnaka}

\institute{
Universidad Cat\'{o}lica del Norte, Instituto de Astronom\'{i}a, 
Avenida Angamos 0610, Antofagasta, Chile\\
\email{k1.ohnaka@gmail.com}
\and
Max-Planck-Institut f\"{u}r Radioastronomie, 
Auf dem H\"{u}gel 69, 53121 Bonn, Germany
}

\date{Received / Accepted }

\abstract
{
 Dust formation is thought to play an important role in the mass loss 
from stars at the asymptotic giant branch (AGB); however,   
where and how dust forms  is still open to debate.
}
{
We present visible polarimetric imaging observations of the well-studied AGB 
star \whya\ taken with VLT/SPHERE-ZIMPOL as well as high spectral 
resolution long-baseline interferometric observations taken with the AMBER 
instrument at the Very Large Telescope Interferometer (VLTI).  
Our goal is to spatially resolve the dust and molecule formation 
region within a few stellar radii. 
}
{
We observed \whya\ with VLT/SPHERE-ZIMPOL at three wavelengths in the 
continuum (645, 748, and 820~nm), in the H$\alpha$ line at 656.3~nm, and 
in the TiO band at 717~nm.  
The VLTI/AMBER observations were carried out in the wavelength region 
of the CO first overtone lines near 2.3~\micron\ with a spectral 
resolution of 12000. 
}
{
Taking advantage of the polarimetric imaging capability of SPHERE-ZIMPOL 
combined with the superb adaptive optics performance, 
we succeeded in spatially resolving three clumpy dust clouds 
located at $\sim$50~mas ($\sim$2~\RSTAR) 
from the central star, revealing dust formation very close to the star.  
The AMBER data in the individual CO lines suggest a molecular 
outer atmosphere extending to $\sim$3~\RSTAR.  
Furthermore, the SPHERE-ZIMPOL image taken over the \Ha\ line shows emission 
with a radius of up to $\sim$160~mas ($\sim$7~\RSTAR).  We found that 
dust, molecular gas, and \Ha-emitting hot gas coexist within 2--3~\RSTAR.  
Our modeling suggests that the observed polarized intensity maps can 
reasonably be explained by large (0.4--0.5~\micron) grains of 
\corundum,\ \forsterite,\ or \enstatite\ in an optically thin shell 
($\tau_{\rm 550nm} = 0.1\pm0.02$) 
with an inner and outer boundary radius of 1.9--2.0~\RSTAR\ and 
$3\pm0.5$~\RSTAR, respectively. 
The observed clumpy structure can be reproduced by a density 
enhancement of a factor of $4\pm1$.  
}
{
The grain size derived from our modeling of the SPHERE-ZIMPOL polarimetric 
images is consistent with the prediction of the hydrodynamical models 
for the mass loss driven by the scattering due to micron-sized grains.  
The detection of the clumpy dust clouds close to the star lends support 
to the dust formation induced by pulsation and large convective 
cells as predicted by the 3-D simulations for AGB stars. 
}

\keywords{
techniques: polarimetric -- 
techniques: interferometric -- 
stars: imaging -- 
stars: AGB and post-AGB -- 
(stars:) circumstellar matter --
stars: individual: W~Hya
}   

\titlerunning{Clumpy dust formation and molecular outer atmosphere 
of the AGB star W~Hya
}
\authorrunning{Ohnaka et al.}
\maketitle

\section{Introduction}
\label{sect_intro}

Low- and intermediate-mass stars experience significant mass loss at 
late stages of their evolution, particularly on 
the asymptotic giant branch (AGB).  
It is often argued that the levitation of the atmosphere by 
the large-amplitude stellar pulsation leads to dust formation, and 
the radiation pressure that the dust grains receive by absorbing stellar 
photons can initiate mass outflows.  
However, in the case of oxygen-rich AGB stars, 
the hydrodynamical simulations of Woitke (\cite{woitke06}) and 
H\"ofner (\cite{hoefner07}) show that the radiation pressure on dust grains 
caused by the absorption of stellar photons is not sufficient to drive 
the mass loss with the mass-loss rates observed in these stars.  
The reason is that the iron-rich silicate, which efficiently absorbs 
stellar photons in the visible and near-IR and is therefore  crucial 
for driving the mass loss, cannot exist in the vicinity of the star 
because the dust temperature exceeds the sublimation temperature. 
On the other hand, \corundum\ and the iron-poor silicate can exist 
much closer to the star because their opacity in the visible and near-IR is 
low.  However, and exactly for this reason, it does not absorb stellar photons 
sufficiently to drive mass outflows.  
To solve this problem, H\"ofner (\cite{hoefner08}) proposes that the 
radiation pressure on micron-sized iron-free silicate grains due to 
scattering---and not to absorption---can drive outflows in oxygen-rich AGB 
stars.

High angular resolution observations have been revealing the presence of 
dust in the vicinity of the central star.  
The visible interferometric observations of Miras and semi-regular variables 
by Ireland et al. (\cite{ireland04}) show an increase in the angular size 
toward shorter wavelengths, which is attributed to the increase in scattered 
light by dust in the inner circumstellar envelope.  
The long-baseline polarimetric interferometry of the Mira stars R~Car 
and RR~Sco carried out by Ireland et al. (\cite{ireland05}) suggests the 
presence of transparent grains within 3~\RSTAR.  
More recently, \mbox{Norris} et al. (\cite{norris12}) have carried out 
polarimetric interferometry at 1.04--2.06~\micron\ 
for three AGB stars (including \whya\ presented in this paper) using the 
the aperture-masking technique with the NACO instrument of VLT. 
They measured the visibilities (i.e., amplitude of the Fourier transform 
of the intensity distribution of the object) in two perpendicular polarization 
directions.  
Because the visibility is sensitive to the size and shape of the object, 
the ratio of the visibilities measured in two polarization directions allowed 
them to detect scattered light from a dust shell very close to the star 
at 1.6--2~\RSTAR. 
Furthermore, they derived a grain size of $\sim$0.3~\micron, in agreement 
with the theory of H\"ofner (\cite{hoefner08}).  
\mbox{Norris} et al. (\cite{norris12}) suggest that iron-free silicates 
such as forsterite (\forsterite) and enstatite (\enstatite) or corundum 
(\corundum) should be responsible for the scattering.  
The mid-IR interferometric observations of the semi-regular M giant RT~Vir 
with the MIDI instrument at VLTI also lend support to the presence of 
iron-free silicate between 2 and 3~\RSTAR\ (Sacuto et al. \cite{sacuto13}).

The red giant \whya\ is one of the well-studied oxygen-rich AGB stars.  
Thanks to its brightness and proximity ($78^{+6.5}_{-5.6}$~pc, 
Knapp et al. \cite{knapp03}), 
it has been studied with various observational techniques 
from the visible to the radio.   While it is classified 
as a semi-regular variable with spectral types of M7.5--9 on Simbad, 
its light curve shows clear periodicity 
(e.g., Woodruff et al. \cite{woodruff08}) 
with a period of 389~days (Uttenthaler et al. \cite{uttenthaler11}), 
although the variability amplitude of $\Delta V \approx 3$ is noticeably 
smaller than that of typical Mira stars.  
Therefore, it is often treated as a Mira star in the literature 
(see Uttenthaler et al. \cite{uttenthaler11} for a discussion of the 
classification).

Infrared interferometric observations of \whya\ reveal an extended 
atmosphere.  
The uniform-disk diameter derived between 1.1 and 3.8~\micron\ 
by Woodruff et al. (\cite{woodruff08}, \cite{woodruff09}) shows that 
the angular size in the wavelength regions relatively free of 
 \HOH\ bands is 30--38~mas, while the angular size 
increases up to $\sim$70~mas in the \HOH\ bands, 
suggesting the presence of an \HOH\ shell.  
The mid-IR interferometric observations of 
Zhao-Geisler et al. (\cite{zhao-geisler11}) with VLTI/MIDI show 
that the angular diameter is 75--80~mas at 8--10~\micron\ and increases 
to 100--105~mas from 10 to 13~\micron.  These authors interpret that 
the angular diameters at 8--10~\micron\ represent the size of the \HOH\ 
shell, while the increase in the angular diameter longward of 10~\micron\ 
can be attributed to the dust envelope. 

Khouri et al. (\cite{khouri15}) present a detailed modeling of the dust 
envelope of \whya\  using the spectral energy distribution (SED) and the dust 
spectral features observed from the near-IR to the sub-mm domain, as well as 
the aforementioned high spatial resolution observations of Norris et al. 
(\cite{norris12}) and Zhao-Geisler et al. (\cite{zhao-geisler11}).  
The Khouri et al. model consists of a gravitationally bound shell of large 
(0.3~\micron) grains of \corundum\ or \forsterite\ between 1.7 and 
2.0~\RSTAR\ 
and an outer amorphous silicate shell with an inner radius of $>$25~\RSTAR.  
They derived a current dust mass-loss rate of $4\times10^{-10}$~\MSOLPERYR\ 
and suggest that there has been a change in the mass-loss rate of a factor 
of 2--3 within the last 4500~years. 

In this paper, we present visible polarimetric imaging observations 
of \whya\ with the VLT/SPHERE-ZIMPOL instrument as well as high spatial 
and high spectral resolution near-IR interferometric observations with 
the VLTI/AMBER instrument.  Our goal is to probe the dust and 
molecular gas close to the central star.

\begin{table*}
\begin{center}
\caption {
Summary of the SPHERE-ZIMPOL observations of \whya. 
DIT: Detector integration time.  
NDIT: Number of frames.
N$_{\rm exp}$: Number of exposures of each polarization component at each 
dithering position. 
N$_{\rm dither}$: Number of dithering positions. 
AM: Airmass. 
The Strehl ratios in the visible are computed from the $H$-band Strehl ratios 
for \whya, while they were measured from the observed ZIMPOL images for 
HD121653. 
}

\begin{tabular}{l c c c c c l l l l l}\hline
\# & $t_{\rm obs}$ & DIT  & NDIT & N$_{\rm exp}$ & N$_{\rm dither}$ & Filter 
& Seeing & AM & Strehl & Strehl \\
   & (UTC)       & (sec)&      &               &              & (ND/cam1/cam2) 
& (\arcsec) &  &  ($H$) & (visible)\\
\hline
\multicolumn{11}{c}{\whya : 2015 July 08 (UTC)}\\
\hline
1 & 00:23:22 & 10  & 10 & 1 & 3 &  ---/CntHa/NHa & 0.94 & 1.04 & 0.86 & 0.37\\
2 & 01:09:06 & 1.2 & 10 & 8 & 3 &  ---/TiO717/Cnt748 & 1.08 & 1.11 &0.88&0.51 \\
3 & 01:47:22 & 2   & 6  & 6 & 3 &  ND1/Cnt820/Cnt820 & 1.03 & 1.20 &0.79&0.39 \\
\hline
\multicolumn{11}{c}{HD121653: 2015 July 08 (UTC)}\\
\hline
C1 & 02:28:20 & 20 & 4 & 1 & 3 &  ---/CntHa/NHa & 1.27 & 1.31 &0.56&0.030\\
C2 & 02:54:50 & 10 & 2 & 4 & 3 &  ---/TiO717/Cnt748 & 1.20 & 1.45 &0.59&0.071\\
C3 & 03:16:42 & 10 & 2 & 4 & 3 &  ---/Cnt820/Cnt820 & 1.26 & 1.59 &0.57&0.082\\
\hline
\label{obs_log_sphere}
\vspace*{-7mm}

\end{tabular}
\end{center}
\end{table*}

\begin{figure*}
\sidecaption
\rotatebox{0}{\includegraphics[width=12cm]{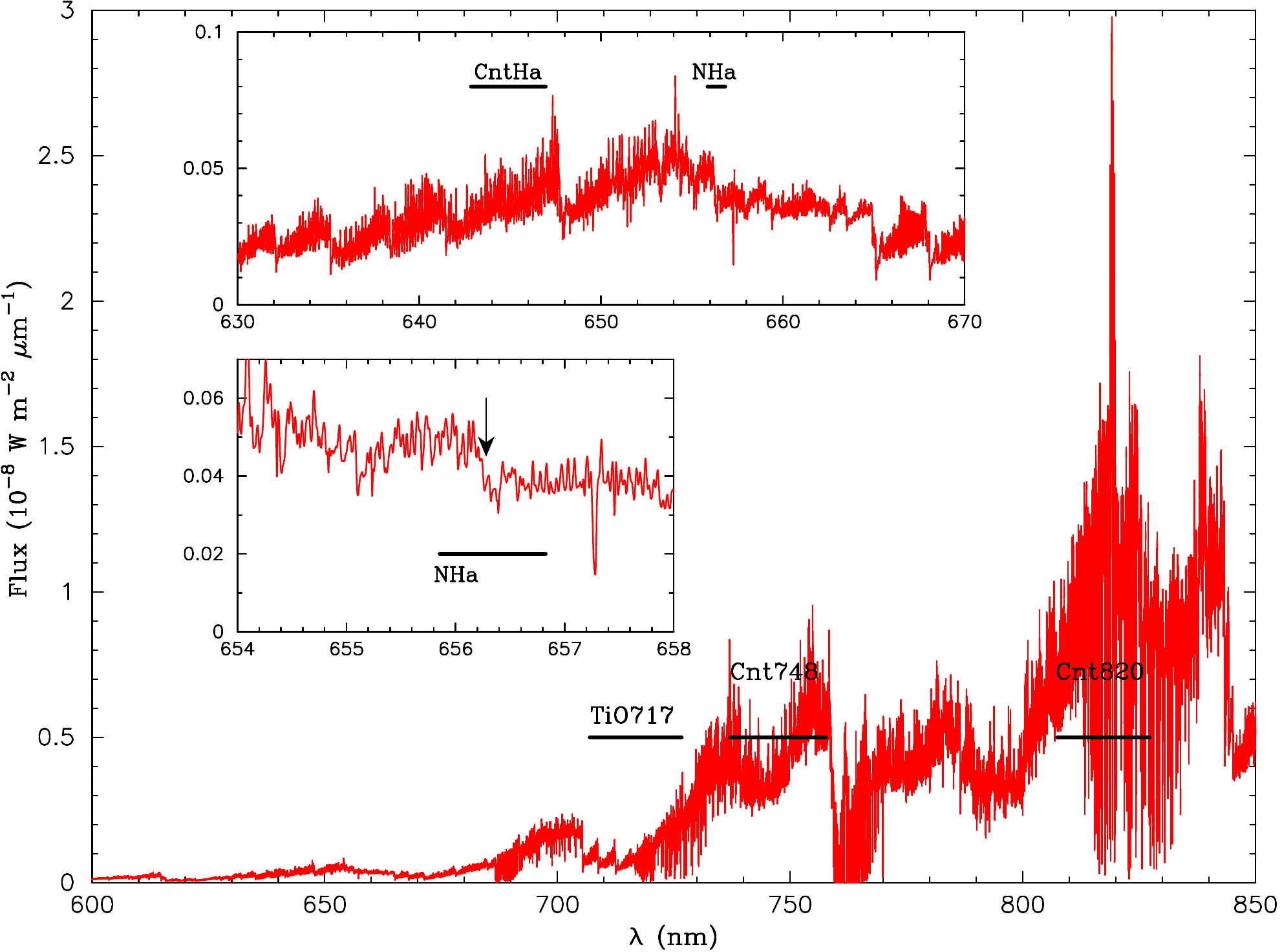}}
\caption{
High resolution visible spectrum of \whya. 
The spectrum is based on the data obtained by Uttenthaler et al. 
(\cite{uttenthaler11}), and we applied the flux calibration to their spectrum 
as described in Sect.~\ref{subsect_obs_zimpol}.  
The FWHMs of five filters used in our SPHERE-ZIMPOL 
observations are marked with the horizontal bars.  The insets show 
enlarged views of the spectral region of the CntHa and NHa filters. 
The arrow in the lower inset marks the position of the \Ha\ line. 
}
\label{whya_vis_spectrum}
\end{figure*}

\section{Observations}
\label{sect_obs}

\subsection{SPHERE-ZIMPOL polarimetric imaging observations}
\label{subsect_obs_zimpol}

VLT/SPHERE is a high spatial resolution and high contrast imaging 
instrument mounted on the Unit Telescope (UT) 3; it is 
equipped with an extreme adaptive optics (AO) system for 
0.55 to 2.32~\micron\ (Beuzit et al. \cite{beuzit08}).  
The ZIMPOL instrument is a module for nearly diffraction-limited 
polarimetric imaging (as well as classical, non-polarimetric 
imaging) at 550--900~nm (Thalmann et al. \cite{thalmann08}). 
Our SPHERE-ZIMPOL observations of \whya\ 
(Program ID: 095.D-0397, P.I.: K.~Ohnaka) took place 
on 2015 July 08 (UTC) in P2 mode, in which 
the field orientation remained fixed.  
The K3III star HD121653 ($V$ = 7.2) was observed as a reference of the point 
spread function (PSF). 
The angular diameter of HD121653 is $0.789\pm 0.022$~mas 
(CalVin database\footnote{http://www.eso.org/observing/etc/bin/gen/\\
form?INS.NAME=CALVIN+INS.MODE=CFP}), 
which appears as a point source with the spatial resolution of SPHERE-ZIMPOL.  
The $V$ magnitude of \whya\ at the time of our SPHERE-ZIMPOL 
observations is estimated to be $\sim$7, corresponding to phase 0.9 
(pre-maximum light) of the light curve of the 
American Association of Variable Star Observers (AAVSO).  
The summary of our SPHERE-ZIMPOL observations is given in 
Table~\ref{obs_log_sphere}. 

We used five filters: CntHa (central wavelength \wlc\ = 644.9~nm, 
FWHM = 4.1~nm), NHa (\wlc\ = 656.34~nm, FWHM = 0.97~nm), 
TiO717 (\wlc\ = 716.8~nm, FWHM = 19.7~nm), Cnt748 (\wlc\ = 747.4~nm, 
FWHM = 20.6~nm), and Cnt820 (\wlc\ = 817.3~nm, FWHM = 19.8~nm). 
Figure~\ref{whya_vis_spectrum} shows the high  resolution spectrum 
of \whya\ obtained by Uttenthaler et al. (\cite{uttenthaler11}) 
at nearly the same variability phase as our SPHERE-ZIMPOL observations 
(photometric calibration of the spectrum carried out  
as described below).  
The visible spectrum of \whya\ is dominated by the prominent TiO bands.  
As the figure shows, three filters, CntHa, Cnt748, 
and Cnt820, sample the \mbox{(pseudo-)}continuum regions that are relatively 
free of TiO bands.  
The TiO717 filter samples the wavelength region of the TiO band. 
The NHa filter covers the wavelength region of the \Ha\ line, as shown in 
the inset.  

The SPHERE-ZIMPOL instrument, which is equipped with two cameras 
(cam1 and cam2), enables us to observe a given target with the same or 
different filters simultaneously.  We observed \whya\ 
by using the filter pairs of (CntHa, NHa), (TiO717, Cnt748), and (Cnt820, 
Cnt820) for cam1 and cam2.  
For the observations of \whya\ with the Cnt820 filter, the neutral 
density filter ND1 was inserted in the path common to cam1 and cam2.  
The pixel scale of the two cameras is 3.628~mas.  
For each target and with each filter pair, we took $N_{\rm exp}$ exposures 
for each of the Stokes \Qplus, \Qminus, \Uplus, and \Uminus\ components, 
with NDIT frames in each exposure.  We repeated this procedure at three 
different dithering positions.  

The SPHERE instrument records the $H$-band Strehl ratios 
in separate FITS files\footnote{Classified as ``OBJECT, AO'' 
in the ESO data archive.}.  
As listed in Table~\ref{obs_log_sphere}, 
the median $H$-band Strehl ratio during the observations of \whya\ was 
0.79--0.88, which corresponds to Strehl ratios of 0.37--0.51 at the 
wavelengths of the ZIMPOL observations, 
if we assume that the Fried parameter is proportional to $\lambda^{6/5}$.  
However, the median Strehl ratio during the observations of the PSF reference 
HD121653 was significantly lower: 0.56--0.59 in the $H$ band.  
The Strehl ratios in the visible derived from the observed images of 
HD121653 are 0.030--0.082 at the wavelengths of the ZIMPOL observations, 
which are 5--12 times lower than those for \whya.  
The reason for the lower Strehl ratios for the PSF reference is likely the 
worse seeing during the observations of HD121653 than for \whya.  
Because of this significant difference in the AO performance between \whya\ 
and the PSF reference, we did not deconvolve the images of \whya.

We reduced the raw data using the SPHERE pipeline
version~0.15.0-2\footnote{Available at
  ftp://ftp.eso.org/pub/dfs/pipelines/sphere}.  
Each exposure was processed with the reduction pipeline, which produces 
the image of the \Qplus\ or \Qminus\ or \Uplus\ or \Uminus\ component 
as well as the intensity of each component (\IQplus, \IQminus, \IUplus, and 
\IUminus) for each camera.  The output images were aligned and 
added to produce the average images of the polarization components and their 
associated intensity.  Then the Stokes parameters $I$, $Q$, and $U$ were 
computed as 
\begin{eqnarray}
Q  = \frac{\Qplus - \Qminus}{2}, \,\, 
U = \frac{\Uplus - \Uminus}{2}, \,\,
I_Q = \frac{\Qplus + \Qminus}{2}, \,\,
I_U = \frac{\Uplus + \Uminus}{2}, \nonumber
\end{eqnarray}
\[
I = \frac{I_Q + I_U}{2}. 
\]
The polarized intensity \IP\ and the degree of linear polarization \PL\ 
as well as the position angle $\theta$ of the polarization vector were 
calculated from the Stokes parameters as follows:
\[
\IP = \sqrt{Q^2+U^2}, \,\,\, \PL = \IP / I, \,\,\, \theta = \frac{1}{2} \arctan(U/Q). 
\]
The orientation of the images was set so that north is up and  east to the left, 
based on the following information kindly provided by M. van den Ancker 
in the User Support Department of ESO: 
the images on cam1 were up-down flipped, while those on cam2 were rotated by 
180\degr.  Since the position angle of the y-axis of the detector is 
357.95\degr $\pm$ 0.55\degr, the images (both cam1 and cam2) were rotated 
clockwise by 2.05\degr.

The flux of our PSF reference star HD121653 with the filters used in 
our observations is not known, which means that we cannot use this star for 
the flux calibration of the SPHERE-ZIMPOL intensity maps of \whya.  
Therefore, we carried out the flux calibration of the intensity maps using 
the high resolution visible spectrum of \whya\ presented in Uttenthaler et al. 
(\cite{uttenthaler11}), which was obtained approximately at the same 
variability phase (although in a different variability cycle).  
We first scaled the spectrum so that the flux obtained with the 
$V$-band filter matches the $V$ magnitude of 7 at the time of our 
SPHERE observations estimated from the AAVSO light curve.  
From this flux-calibrated visible spectrum of \whya, 
we computed the flux for each filter used in our 
observations.  
We approximated the transmission curves of the filters 
with a top-hat function specified with the central wavelength and FWHM.  
The resulting fluxes with five filters are given in 
Table~\ref{whya_photometry}.  
The SPHERE-ZIMPOL intensity maps were scaled so that the flux integrated 
within a radius of 1\farcs4 matches these fluxes.  
We chose this radius, because we needed to set the radius to be 
as large as possible to include the whole detected flux 
and still stay within the detector.

\begin{table}
\begin{center}
\caption {
 Flux derived for the five filters of our SPHERE-ZIMPOL observations 
of \whya.  These values are used for the flux calibration of the SPHERE-ZIMPOL 
images. 
}

\begin{tabular}{l l c}\hline
Filter & $\lambda_{\rm c}$ (nm) &  Flux (W~m$^{-2}$~$\micron^{-1}$) \\
\hline
CntHa  & 644.9  & $3.78\times10^{-10}$ \\
NHa    & 656.34 & $4.33\times10^{-10}$ \\
TiO717 & 716.8  & $1.09\times10^{-9}$ \\
Cnt748 & 747.4  & $4.36\times10^{-9}$ \\
Cnt820 & 817.3  & $1.05\times10^{-8}$ \\
\hline
\label{whya_photometry}
\vspace*{-7mm}

\end{tabular}
\end{center}
\end{table}

\begin{table*}
\begin{center}
\caption {
Summary of the VLTI/AMBER observations of \whya\ and the calibrators. 
$B_{\rm p}$: Projected baseline length.  PA: Position angle of the baseline 
vector projected onto the sky. 
DIT: Detector Integration Time.  $N_{\rm f}$: Number of frames in each 
exposure.  $N_{\rm exp}$: Number of exposures. 
The seeing and the coherence time ($\tau_0$) were measured in the visible. 
}

\begin{tabular}{l c c c c r l }\hline
\# & $t_{\rm obs}$ & $B_{\rm p}$ & PA     & Seeing   & $\tau_0$ &
${\rm DIT}\times{\rm N}_{\rm f}\times{\rm N}_{\rm exp}$ \\ 
& (UTC)       & (m)       & (\degr) & (\arcsec)       &  (ms)    &  (ms)  \\
\hline
\multicolumn{7}{c}{\whya: 2014 April 22 (UTC)}\\
\hline
1 & 00:33:58 & 7.02/11.23/12.37 & 93/$-5$/29    & 1.18 & 4.3 & $120\times500\times1$ \\
2 & 00:36:05 & 7.09/11.23/12.41 & 93/$-5$/29    & 1.02 & 5.0 & $120\times500\times1$ \\
3 & 00:38:12 & 7.16/11.23/12.45 & 93/$-5$/30    & 0.91 & 5.6 & $120\times500\times1$ \\
4 & 00:40:19 & 7.23/11.23/12.49 & 93/$-5$/30    & 0.84 & 6.1 & $120\times500\times1$ \\
5 & 00:42:25 & 7.31/11.23/12.54 & 94/$-4$/31    & 0.99 & 5.1 & $120\times500\times1$ \\
6 & 01:11:44 & 8.23/11.22/13.15 & 96/0/38       & 0.87 & 5.8 & $120\times500\times1$ \\
7 & 01:13:50 & 8.29/11.22/13.19 & 97/0/39    & 1.07 & 4.7 & $120\times500\times1$ \\
8 & 01:15:57 & 8.35/11.22/13.24 & 97/1/39    & 0.95 & 5.3 & $120\times500\times1$ \\
9 & 01:18:05 & 8.41/11.22/13.29 & 97/1/40    & 0.73 & 6.9 & $120\times500\times1$ \\
10 & 01:20:12 & 8.47/11.22/13.33 & 97/1/40   & 0.79 & 6.4 & $120\times500\times1$ \\
\hline
\multicolumn{7}{c}{\alfcenA: 2014 April 22 (UTC)}\\
\hline
C1 & 00:13:10 & 7.48/10.03/15.66 & 37/$-17$/6 & 0.69& 7.4 &$120\times500\times5$\\
C2 & 00:53:34 & 7.97/10.13/15.71 & 50/$-10$/16 & 0.97& 5.3&$120\times500\times5$\\
\hline
\multicolumn{7}{c}{Canopus: 2014 April 21 (UTC)}\\
\hline
C3 & 23:54:02 & 11.20/8.42/13.15 & 148/51/108 & 0.92 & 5.6&$120\times500\times5$\\
\hline
\label{obs_log_amber}
\vspace*{-7mm}

\end{tabular}
\end{center}
\end{table*}

\subsection{High spectral resolution VLTI/AMBER observations}
\label{subsect_obs_amber}

The near-IR VLTI instrument AMBER (Petrov et al. \cite{petrov07}), 
which operates at 1.3--2.4~\micron, 
combines three UTs or 1.8~m Auxiliary Telescopes (ATs) and 
allows us to achieve a spatial resolution of 3~mas (at 2~\micron) with 
the current maximum baseline of 140~m.  The AMBER instrument is equipped 
with three spectral resolutions, 35, 1500, and 12000.  With the highest 
spectral resolution of 12000, it is possible to resolve individual atomic 
and molecular lines.  
The interferometric observables measured with AMBER are visibility, 
closure phase (CP), and differential phase (DP).  
Visibility is the amplitude of the Fourier transform of the object's intensity 
distribution on the sky and contains information about the size and shape 
of the object.  The CP is the sum of the Fourier phases on three 
baselines around a triangle formed by three telescopes.  
Deviations of CP from 0 or 180\degr\ indicate asymmetry of the object.  
The DP represents the photocenter shift in spectral features with respect to 
the continuum.  
The AMBER instrument also records the spectrum of the same wavelength 
region simultaneously with the interferometric fringes. 

We observed \whya\ with VLTI/AMBER on 2014 April 22 (UTC) using the 
AT configuration A1-B2-C1, which covered projected baseline lengths 
from 7.0 to 13.3~m (Program ID: 093.D-0468, P.I.: K.~Ohnaka).  
A summary of our AMBER observations is given in Table~\ref{obs_log_amber}. 
The data taken at projected baselines shorter than 13.3~m 
correspond to the first visibility lobe of \whya\ in the continuum (see 
Fig.~\ref{whya_amber_lddfit}c).  Therefore, these data are appropriate 
for obtaining approximate sizes of the star and the extended atmosphere.  
The data with the projected baseline lengths shorter than the UT 
aperture of 8~m can be obtained with AO instruments, by speckle
interferometry, or by aperture-masking and, if taken simultaneously, 
would be complementary to our AMBER data.  However, with the absence of such 
single-dish data, the AMBER data points with projected baselines shorter than 
8~m are important for constraining the size of the star.  
The variability phase at the time of our AMBER observations is estimated 
to be 0.77 from the AAVSO light curve (but in a different variability cycle).  
This is relatively close to the phase at the time of our SPHERE-ZIMPOL 
observations and allows us to measure the size of the dust shell 
in terms of the stellar radius. 
The wavelength region between 2.28 and 2.31~\micron\ near the CO first 
overtone 2--0 band head was observed with the spectral resolution 
of 12000.  Thanks to the high brightness of \whya\ ($K \approx -3$, 
estimated from the $K$-band light curve of 
Whitelock et al. \cite{whitelock00}), it was possible to achieve reasonable 
fringe S/N without the fringe tracker FINITO (\whya\ saturates FINITO 
in the $H$ band).  We observed \alfcenA\ (G2V, $K = -1.5$, uniform-disk 
diameter = $8.314 \pm 0.016$~mas, Kervella et al. \cite{kervella03}) and 
Canopus ($\alpha$~Car, F0II) as an interferometric and spectroscopic 
calibrator, respectively. 

The recorded fringes were processed with the amdlib 
version~3.0.8\footnote{Available at 
http://www.jmmc.fr/data\_processing\_amber.htm}, 
which is based 
on the P2VM algorithm (Tatulli et al. \cite{tatulli07}; 
Chelli et al. \cite{chelli09}).  
Details of the reduction are described in Ohnaka et al. (\cite{ohnaka09}, 
\cite{ohnaka11}, and \cite{ohnaka13}). 
We checked for a systematic difference in the calibrated visibilities, CPs, 
and DPs by selecting the best 20\% and 80\% frames in terms of the fringe S/N.  
Since we did not detect any noticeable difference in the results, 
we took the best 80\% of the frames, because the errors are smaller.  
The wavelength calibration and the spectroscopic calibration of the \whya\ 
data were carried out with the method described in 
Ohnaka et al. (\cite{ohnaka09}).

\begin{figure*}[!hbt]
\begin{center}
\resizebox{15.cm}{!}{\rotatebox{0}{\includegraphics{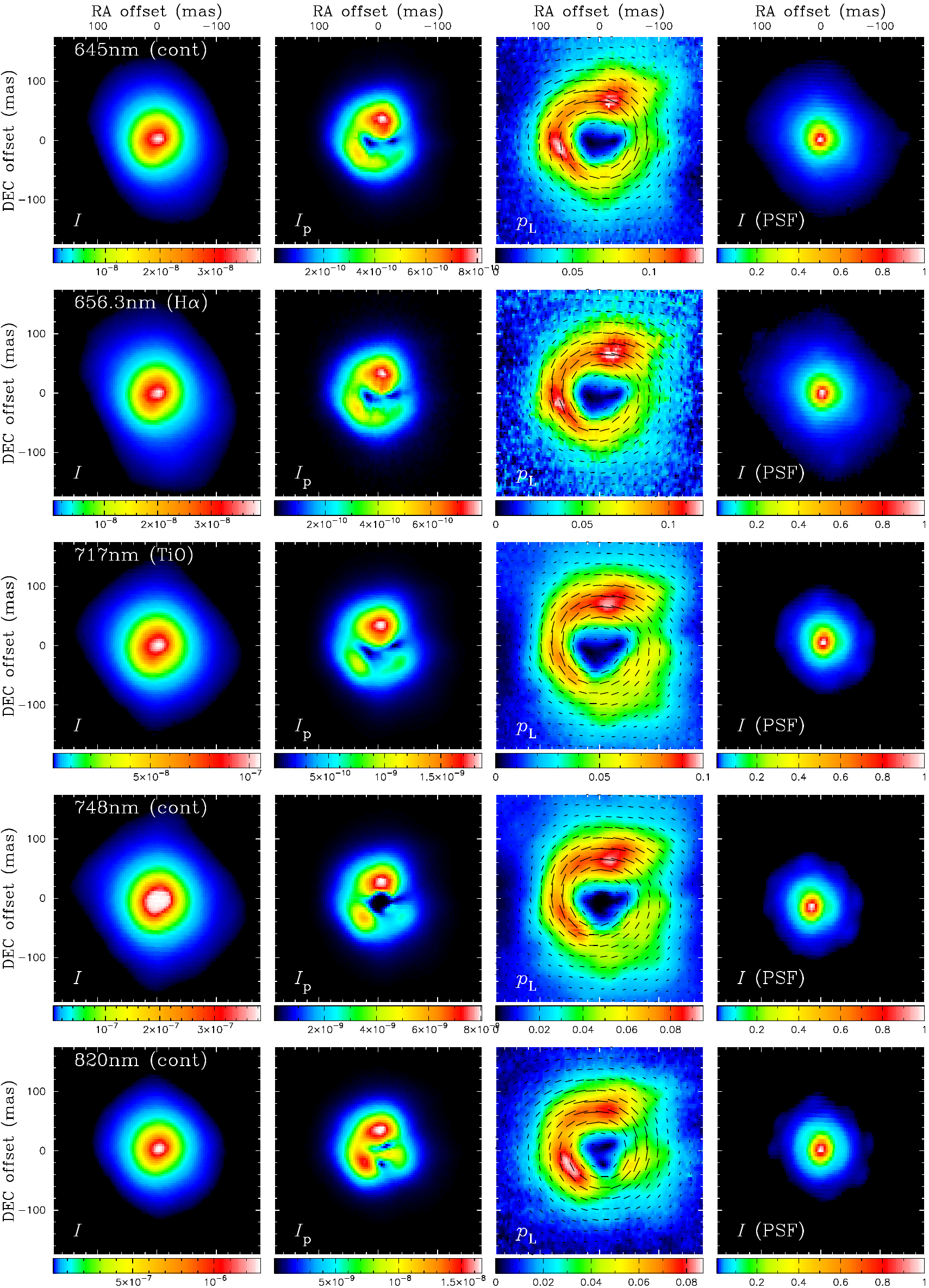}}}
\end{center}
\caption{
Polarimetric imaging observations of \whya\ with SPHERE-ZIMPOL. 
Each row shows the intensity (first column), polarized intensity 
(second column), degree of linear polarization with the polarization vector 
maps overlaid (third column), 
and the intensity of the PSF reference star HD121653 (fourth column). 
The observed images at 645~nm (CntHa, continuum), 656.3~nm (NHa, \Ha), 
717~nm (TiO717, TiO band), 748~nm (Cnt748, continuum), and 820~nm (Cnt820, 
continuum) are shown from  top to  bottom. 
North is up and east to the left in all panels.  
The intensity maps and polarized intensity maps of \whya, which are 
flux-calibrated as described in Sect.~\ref{subsect_obs_zimpol}, are 
shown in units of W~m$^{-2}$~\micron $^{-1}$~arcsec$^{-2}$.  
The color scale of the intensity maps of \whya\ and HD121653 is 
cut off at 1\% of the intensity peak.  
}
\label{whya_zimpol_images}
\end{figure*}

\begin{figure*}
\sidecaption
\resizebox{12cm}{!}{\rotatebox{0}{\includegraphics{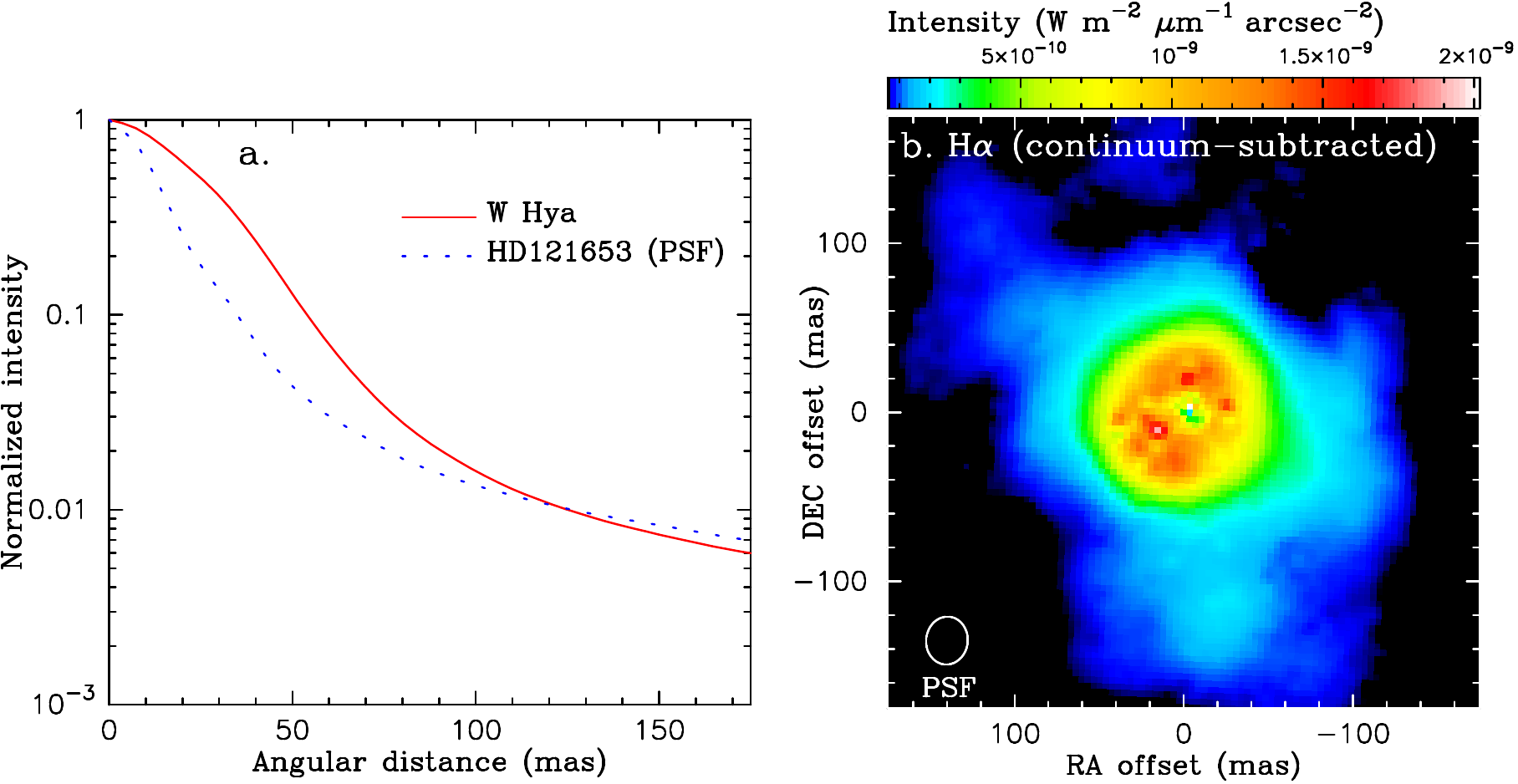}}}
\caption{
{\bf a:} Azimuthally averaged intensity profile of \whya\ (red solid line) and 
the PSF reference HD121653 (blue dashed line) obtained with the CntHa 
filter centered at 645~nm. 
{\bf b:} Continuum-subtracted \Ha\ image of \whya\ derived from the 
SPHERE-ZIMPOL observations.  
}
\label{whya_zimpol_1Davg_Halpha}
\end{figure*}

\section{Results}
\label{sect_res}

\subsection{SPHERE-ZIMPOL polarimetric images}
\label{subsect_res_sphere}

\subsubsection{Clumpy dust clouds}
\label{subsubsect_res_sphere_dust}

Figure~\ref{whya_zimpol_images} shows the intensity $I$, 
polarized intensity \IP, and the degree of linear polarization \PL\ of 
\whya\ observed with five filters, together with the intensity of the 
PSF reference star HD121653.   
The intensity maps of \whya\ (first column of the figure) are noticeably 
more extended than the PSF reference images (fourth column of the figure), 
which means that the 
circumstellar envelope has been spatially resolved.  
This is clearly 
seen in the azimuthally averaged \mbox{1-D} intensity profiles of \whya\ and 
HD121653, which are plotted in Fig.~\ref{whya_zimpol_1Davg_Halpha}a. 
The 2-D Gaussian fit to the observed images of HD121653 results in 
PSF FWHMs of $25\times28$~mas at 645~nm and 656.3~nm, 
$23\times28$~mas at 717~nm and 748~nm, and $24\times30$~mas at 820~nm.

The FWHM of the intensity distribution of \whya\ at 645, 656.3, 717, and 
820~nm is 53, 51, 58, and 46~mas, respectively.  
We note that the central region (within a radius of $\la$20~mas) 
of the image at 748~nm is saturated, which makes is impossible to measure 
the FWHM (and also means that the flux calibration of the Cnt748 image 
is unreliable).  
Haniff et al. (\cite{haniff95}) measured a Gaussian FWHM of 53.6~mas 
at 710~nm (filter FWHM = 10~nm).  
Ireland et al. (\cite{ireland04}) measured the angular size of \whya\ 
from 680 to 940~nm, and 
their FWHMs at the wavelengths of our observations are 
75~mas (717~nm) and 42~mas (820~nm).  
The FWHM of our SPHERE-ZIMPOL image at 717~nm agrees with the result of 
Haniff et al. (\cite{haniff95}), but is much smaller 
than that measured by Ireland et al. (\cite{ireland04}). 
The FWHMs measured at 820~nm by Ireland et al. (\cite{ireland04})  
agree with our values.  
We note that the observations of Ireland et al. (\cite{ireland04}) 
were carried out at phase 0.44 (near minimum light), 
while our observations and the observations of 
Haniff et al. (\cite{haniff95}) took place near maximum light, 
at phase 0.9 and 0.04, respectively. 
The angular size measurements of \whya\ from 1.1 to 3.8~\micron\ 
by Woodruff et al. (\cite{woodruff09}) show that the star appears smaller 
near maximum light than at minimum light.  Therefore, the larger size 
measured by Ireland et al. (\cite{ireland04}) than the present work or 
Haniff et al. (\cite{haniff95}) may be 
due to the difference in the variability phase at the time of the 
observations.

While the intensity maps only show the global extended structure of the 
circumstellar envelope, the polarized intensity maps \IP\ 
(second column in Fig.~\ref{whya_zimpol_images}) reveal more detailed, 
clumpy structures of the innermost region of the envelope. 
We detected a large, bright clump in the 
north of the central star, another large clump in the SE, and a smaller clump 
in the SW. 
These clumps form an incomplete shell with a radius of $\sim$50~mas. 
Moreover, as described in Sect.~\ref{subsect_res_amber}, the angular diameter 
of the central star measured in the continuum with AMBER is 46.6~mas. 
This means that the peak of the dust clumps is found at $\sim$2~\RSTAR, and 
most of the clumps are located within $\sim$3~\RSTAR\ of the 
center of the star.  
In general, 
the polarized intensity represents the map of the optical depth, 
which is proportional to the column density of the scattering dust grains 
in an optically thin case (as presented in Sect.~\ref{sect_modeling}, this is 
the case for \whya).  Therefore, the obtained \IP\ maps indeed reveal 
clumpy dust formation very close to the star, at $\sim$2~\RSTAR.   
We note that since the saturation in the Cnt748 intensity map is limited to 
the central region with a radius of $\sim$20~mas, it does not affect the 
clumpy features in the \IP\ map with Cnt748. 
The radius of the dust formation 
region directly imaged in our SPHERE-ZIMPOL polarimetric imaging is consistent 
with the results of the polarimetric interferometric observations 
of Norris et al. (\cite{norris12}) and the modeling of 11.5~\micron\ 
interferometric data of Wishnow et al. (\cite{wishnow10}).  

Clumpy dust clouds are detected in other nearby AGB stars as well.  
For example, clumpy dust clouds have been imaged in the innermost 
circumstellar envelope of the well-studied carbon-rich AGB stars IRC+10216 
(Weigelt et al. \cite{weigelt98}; Haniff \& Buscher \cite{haniff98}; 
Stewart et al. \cite{stewart16} and references therein) 
and CIT6 (Monnier et al. \cite{monnier00}).  Therefore, the formation 
of clumpy structures might be intrinsic to the mass loss phenomenon in AGB 
stars. 

The errors in the polarized intensity (and in the degree of linear 
polarization as described below) were estimated using the output of the 
SPHERE pipeline.  
The relative errors in the polarized intensity are 3--5\% at 645~nm, 
4--8\% at 656.3~nm, and $\sim$2\% at 717, 748, and 820~nm.  
In the region with very low polarized intensity near the center, which 
appears in dark blue between the clumps in the \IP\ maps in 
Fig.~\ref{whya_zimpol_images}, 
the relative errors amount to 
50--60\% at 645 and 656.3~nm, 40\% at 717~nm, and 30\% at 820~nm 
(since the central region of the \IP\ map at 748~nm is affected by the 
aforementioned saturation issue, we excluded it from the error estimate 
of the central region).

\begin{figure*}
\begin{center}
\resizebox{\hsize}{!}{\rotatebox{0}{\includegraphics{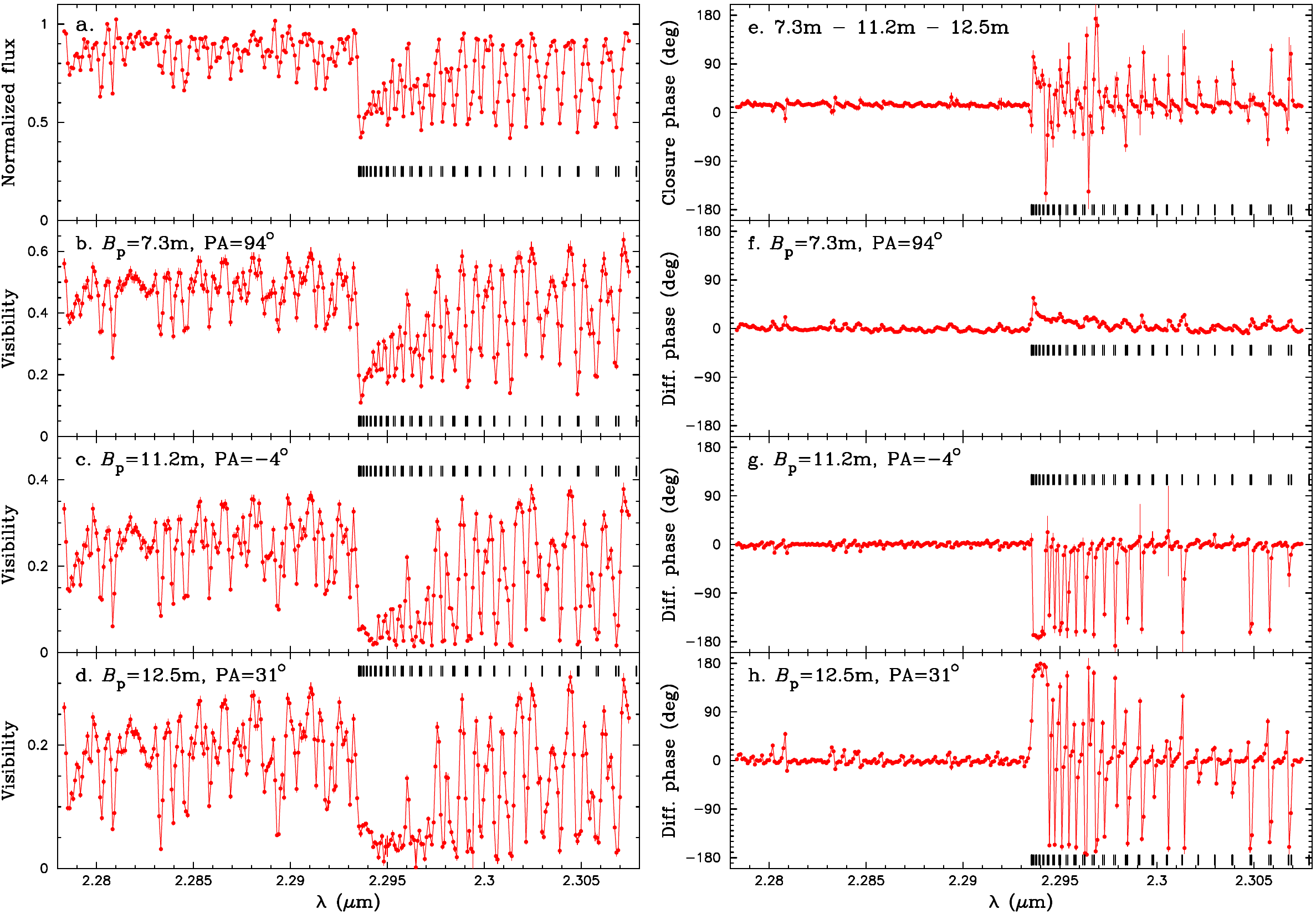}}}
\end{center}
\caption{
VLTI/AMBER observations of \whya\ with a spectral resolution of 12000 
(data set \#5).  
{\bf a:} Observed spectrum. 
{\bf b}--{\bf d:} Visibilities. 
{\bf e:} Closure phase.  
{\bf f}--{\bf h:} Differential phases. 
The positions of the CO lines are indicated by the ticks. 
}
\label{whya_results_amber}
\end{figure*}

The third column of Fig.~\ref{whya_zimpol_images} shows the maps of the degree 
of linear polarization at five wavelengths, with the polarization vector maps 
overlaid. 
The polarization vector maps, which show a concentric pattern, 
confirm the shell-like distribution of the clumps.  
The degree of linear polarization decreases slightly with 
wavelength: the maximum at 645, 656.3, 717, 748, and 820~nm is 13\%, 12\%, 
10\%, 9\%, and 8\%, respectively.  
The absolute errors in the degree of linear polarization in the clumps 
are 0.5\% at 645~nm (i.e., $\PL\ = 13 \pm 0.5$\%), 0.7\% at 656.3~nm, 
0.1\% at 717 and 748~nm, and 0.2\% at 820~nm.  In the central region with 
very low degree of polarization, the absolute errors in \PL\ are 
0.1--0.2\% at 645 and 656.3~nm, 0.03\% at 717~nm, 0.04--0.07\% at 820~nm.

The image of \whya\ in the SO line at 215.2~GHz taken by 
Vlemmings et al. (\cite{vlemmings11}) shows that the redshifted and 
blueshifted components are offset by 0\farcs29 in the N-S direction, 
which the authors interpret as a bipolar outflow or a rotating disk.  
However, there is no signature of a bipolar outflow or a rotating disk in the 
SPHERE-ZIMPOL data probably because the dust formation close to 
the star is driven by large convective cells (see Sect.~\ref{sect_discuss}), 
which may mask the signatures of a bipolar structure or a rotating disk.

\subsubsection{Extended \Ha\ emission}
\label{subsubsect_res_sphere_Ha}

The intensity map taken with the NHa filter including the \Ha\ line 
(Fig.~\ref{whya_zimpol_images}, first column, second row) appears 
to be more extended than the image taken with the CntHa filter sampling 
the nearby continuum (Fig.~\ref{whya_zimpol_images}, first column, first row).  
In order to examine the presence of the \Ha\ emission, we subtracted 
the CntHa image from the NHa image and  both images were flux-calibrated as 
described in Sect.~\ref{subsect_obs_zimpol}.  
Because the observations with the NHa and CntHa filters were carried out 
simultaneously, the performance of AO was the same for both filters. 
The continuum-subtracted \Ha\ image  
shown in Fig.~\ref{whya_zimpol_1Davg_Halpha}b reveals emission with a radius 
of $\sim$100~mas ($\sim$4~\RSTAR);  the emission in the south extends 
up to $\sim$160~mas ($\sim$7~\RSTAR).  

The extended \Ha\ emission of \whya\ is similar to the \Ha\ envelope of the 
red supergiant Betelgeuse extending up to 5~\RSTAR\ imaged by 
Hebden et al. (\cite{hebden87}) and more recently by 
Kervella et al. (\cite{kervella16}).  
While the extended \Ha\ emission in Betelgeuse is thought to originate 
in the hot chromosphere, 
the \Ha\ emission in Mira stars is associated with shocks induced by large 
amplitude stellar pulsation (e.g., Gillet et al. \cite{gillet83}, 
\cite{gillet85}).  Our \Ha\ image of \whya\ reveals the propagation of shocks 
as far as 4--7~\RSTAR. 
The presence of such extended \Ha\ emission may not appear to be consistent 
with the weak \Ha\ absorption seen in the visible spectrum shown in 
Fig.~\ref{whya_vis_spectrum}.  However, this weak absorption can be 
interpreted as a result of the absorption being filled in by the extended 
emission.

\begin{figure*}
\sidecaption
\resizebox{12cm}{!}{\rotatebox{0}{\includegraphics{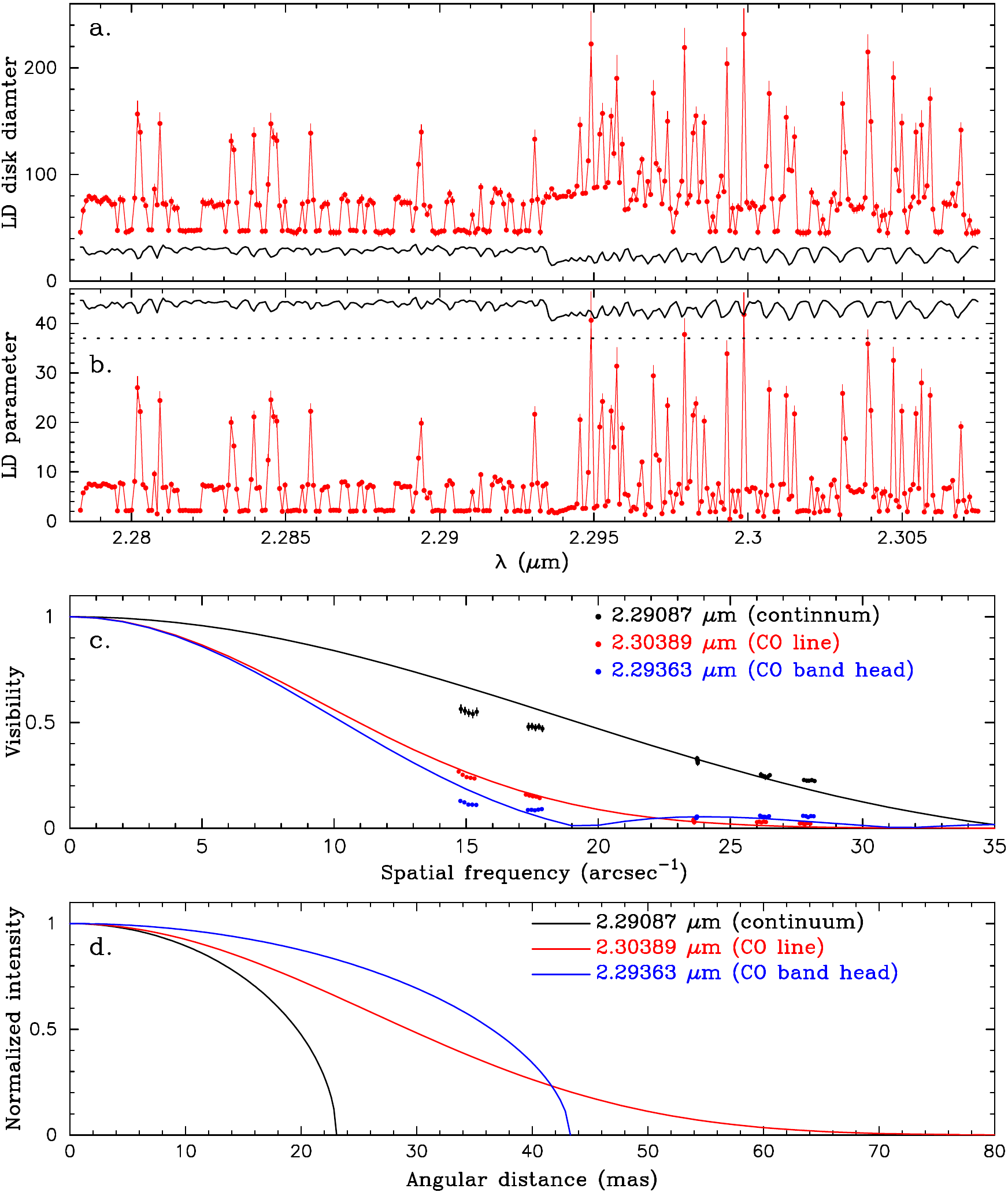}}}
\caption{
Power-law-type limb-darkened disk fit to the AMBER data of \whya.  
{\bf a:} Limb-darkened disk diameter.  The scaled observed spectrum of 
\whya\ is shown by the black solid line. 
{\bf b:} Limb-darkening parameter $\alpha$. 
The black solid line represents the scaled observed spectrum 
with a vertical shift shown by the dashed line. 
{\bf c:} Observed visibilities in the continuum, CO line, and CO band head 
are shown by the dots, while the visibilities from the limb-darkened disk fit 
are plotted by the solid lines. 
{\bf d:} Limb-darkened disk intensity profiles in the continuum, CO line, 
and CO band head. 
}
\label{whya_amber_lddfit}
\end{figure*}

\begin{figure*}[ht]
\begin{center}
\resizebox{\hsize}{!}{\rotatebox{0}{\includegraphics{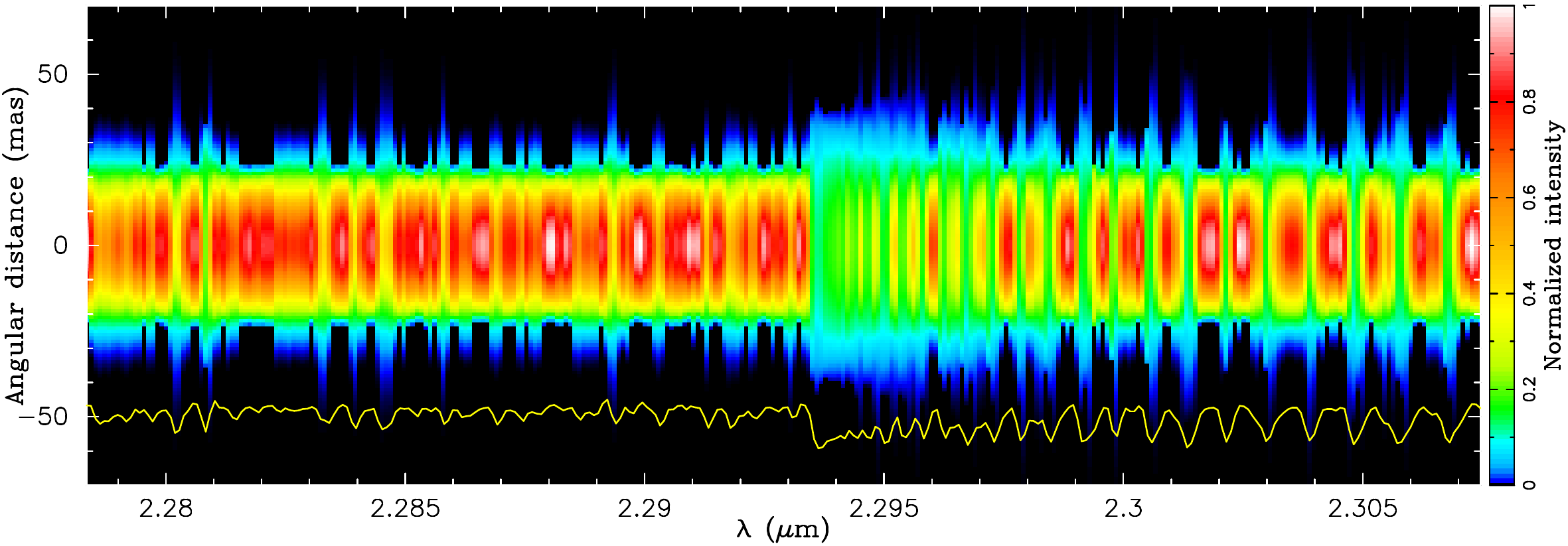}}}
\end{center}
\caption{
Two-dimensional spectrum of \whya\ computed from the limb-darkened disk fit 
shown in Fig.~\ref{whya_amber_lddfit}.  The spatially unresolved spectrum 
is shown in yellow at the bottom. 
}
\label{whya_2Dspec}
\end{figure*}

\subsection{AMBER observations of the central star and molecular outer
  atmosphere}
\label{subsect_res_amber}

Figure~\ref{whya_results_amber} shows the visibilities, differential 
phases, and closure phase of \whya\ observed with VLTI/AMBER from 2.28 to 
2.31~\micron\ (data set \#5).  
In addition to the CO first overtone lines, a number of lines 
are present shortward of the CO band head at 2.2935~\micron.  They are mostly 
\HOH\ and CN lines (e.g., Wallace \& Hinkle \cite{wallace96}).  
To obtain an approximate angular size of the star as a function of wavelength, 
we fitted the observed visibilities with a power-law-type limb-darkened disk 
(Hestroffer et al. \cite{hestroffer97}) in which 
the intensity is described as $I = [1-(p/R)^2]^{\alpha/2}$, where $p$, $R$, 
and $\alpha$ are the impact parameter, limb-darkened disk radius, and 
the limb-darkening parameter ($\alpha = 0$ corresponds to a uniform disk).  

Figure~\ref{whya_amber_lddfit}a shows the obtained limb-darkened disk diameter 
as a function of wavelength.  
On the one hand, the limb-darkened disk angular diameter is 
$46.6\pm0.1$~mas in the continuum and increases up to 230~mas in the CO lines.  
On the other hand, as Fig.~\ref{whya_amber_lddfit}b shows, 
the limb-darkening parameter $\alpha$ is modest in the continuum 
($2.1\pm0.2$; average and standard deviation over the continuum wavelengths), 
but much larger in the CO lines (20--40), suggesting 
significantly stronger limb-darkening in the CO lines. 
In Fig.~\ref{whya_amber_lddfit}d, we plot the limb-darkened disk intensity 
profiles at three representative wavelengths in the continuum, CO line, 
and CO band head.  
The limb-darkening is already remarkable  in the continuum and 
so strong in the CO lines that it leads to Gaussian-like intensity profiles.  

To better visualize the geometrical extension of the star in the continuum 
and in the lines, we generated a \mbox{2-D} spectrum as follows. 
First, the limb-darkened disk intensity profile at each wavelength is 
normalized so that the flux integrated over the stellar disk is equal to the 
observed flux at the corresponding wavelength.  Then, the limb-darkened disk 
intensity profiles are color-coded and placed side by side.  
The resulting 2-D spectrum (Fig.~\ref{whya_2Dspec}) reveals that 
the star shows a halo extending to a radius of 
$\sim$70~mas ($\sim$3~\RSTAR) in the CO lines and to $\sim$35~mas 
($\sim$1.5~\RSTAR) in the weak \HOH\ lines. 

Figure~\ref{whya_amber_lddfit}c shows the fit to the visibilities at three 
wavelengths (continuum, CO bandhead, and CO line).  
The figure suggests that while the limb-darkened disk provides an approximate 
picture of the star's intensity profile, there are noticeable deviations 
from the limb-darkened disk. 
The reduced $\chi^2$ values range from 2 to 100 with a median of 33.2 
for the wavelengths covered by our observations. 
This means that the star is more complex than a symmetric limb-darkened 
disk, possibly with inhomogeneities.  
The detection of non-zero/non-180\degr\ DPs and CPs 
(see Figs.~\ref{whya_results_amber}e--\ref{whya_results_amber}h) 
confirms the presence of such complex structures.  
This makes \whya\ a good target for future aperture-synthesis imaging.

\subsection{Coexistence of dust, CO gas, radio photosphere, SiO masers, 
and \Ha -emitting hot gas}
\label{subsect_res_multicomp}

\begin{figure}
\begin{center}
\resizebox{\hsize}{!}{\rotatebox{0}{\includegraphics{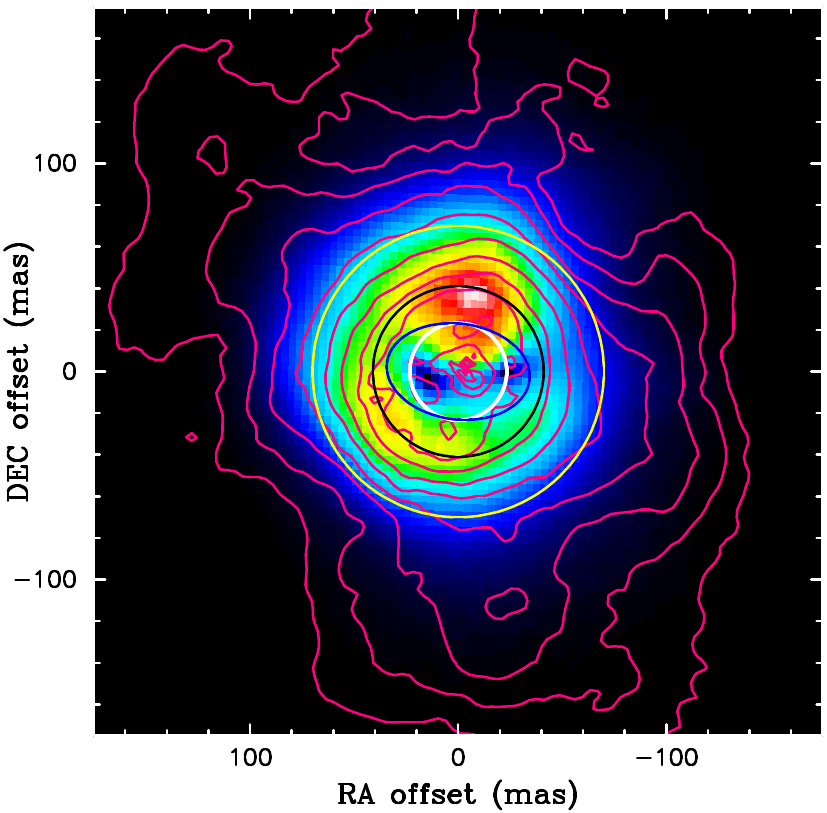}}}
\end{center}
\caption{
Overlay of the distribution of dust (polarized intensity at 645~nm, 
color scale image), hot gas (\Ha\ emission, contours), and the molecular gas 
(extension of the atmosphere derived from the AMBER data in the CO lines, 
outer yellow circle). 
The inner white circle represents the size of the star derived from the AMBER 
data in the continuum. 
The blue ellipse and the middle black circle represent the size of the radio 
photosphere and the SiO maser shell measured at 43~GHz 
by Reid \& Menten (\cite{reid07}), respectively.  
The contours are plotted in logarithmic scale.  The lowest and highest 
contours correspond to 3\% and 100\% of the maximum value, respectively. 
North is up, east to the left.  
}
\label{whya_overlay}
\end{figure}

The extended atmosphere seen in the CO lines overlap with the location of 
clumpy dust clouds detected with SPHERE-ZIMPOL.  
Figure~\ref{whya_overlay} shows the polarized intensity observed at 
645~nm in color scale with the \Ha\ emission represented in the contours. 
The geometrical extension of the atmosphere of $\sim$70~mas ($\sim$3~\RSTAR), 
which is marked with the outer (yellow) circle in the figure, 
encompasses the extension of most of the dust clumps.  
The hot gas associated with the shocks traced with the \Ha\ emission 
extends up to $\sim$7~\RSTAR\ and  overlaps  the distribution of the dust 
and CO gas.

Reid \& Menten (\cite{reid07}) obtained an image of \whya\ in the 43~GHz 
radio continuum with the Karl G. Jansky Very Large Array.  The star shows 
extended emission---the so-called radio photosphere (Reid \& Menten 
\cite{reid97})---that is $69\times46$~mas in size with the major axis 
lying nearly in the E-W direction.  The size of the radio photosphere 
(blue ellipse in Fig.~\ref{whya_overlay}) is much smaller than 
the extension of the atmosphere seen in the 2.3~\micron\ CO lines.  
The variability phase of the radio observations of 
Reid \& Menten (\cite{reid07}) was 0.25, different from the phase of our 
AMBER observations.  While this difference in the variability phase should 
be kept in mind, we also note that the temperature derived from the radio 
observations is rather high, $2380\pm550$~K.  Therefore, the radio continuum 
at 43~GHz may sample the innermost region of the extended atmosphere.  
The SiO $\varv = 1$, $J = 1 - 0$ masers imaged by Reid \& Menten 
(\cite{reid07}) form an incomplete shell with a radius of 41~mas 
(middle black circle), coexisting with the CO gas and dust.  
These multi-wavelength observations suggest the complex, multicomponent 
nature of the outer atmosphere of \whya.

\section{Monte Carlo radiative transfer modeling of the polarized intensity maps}
\label{sect_modeling}

The SPHERE-ZIMPOL polarimetric images provide valuable 
constraints on the properties of the innermost dust envelope, particularly, 
the grain size. 
For this purpose, we used our Monte Carlo radiative transfer code mcmpi\_sim 
(Ohnaka et al. \cite{ohnaka06}), which was also used for the interpretation 
of polarimetric imaging data (Murakawa et al. \cite{murakawa08}).  
The change of the Stokes vector is treated with the scattering matrix 
(see, e.g., Wolf et al. \cite{wolf99}; Gordon et al. \cite{gordon01}), 
whose elements are computed from the complex refractive index of a given 
grain species.  
The output of our Monte Carlo code is the intensity ($I$) and the Stokes 
$Q$ and $U$ images.  To compare these data with the observed values, 
we first convolved 
the model $Q$ and $U$ images, and also the intensity $I$ with the 
observed PSF from the PSF reference star, and then computed the maps of the 
polarized intensity and the degree of linear polarization from the 
convolved $Q$, $U$, and $I$ images.  
We adopted this approach instead of convolving the model \IP\ and 
\PL\ maps because it corresponds to how the \IP\ and 
\PL\ maps were obtained from the observational data.  

\begin{figure*}[hbt]
\begin{center}
\resizebox{\hsize}{!}{\rotatebox{0}{\includegraphics{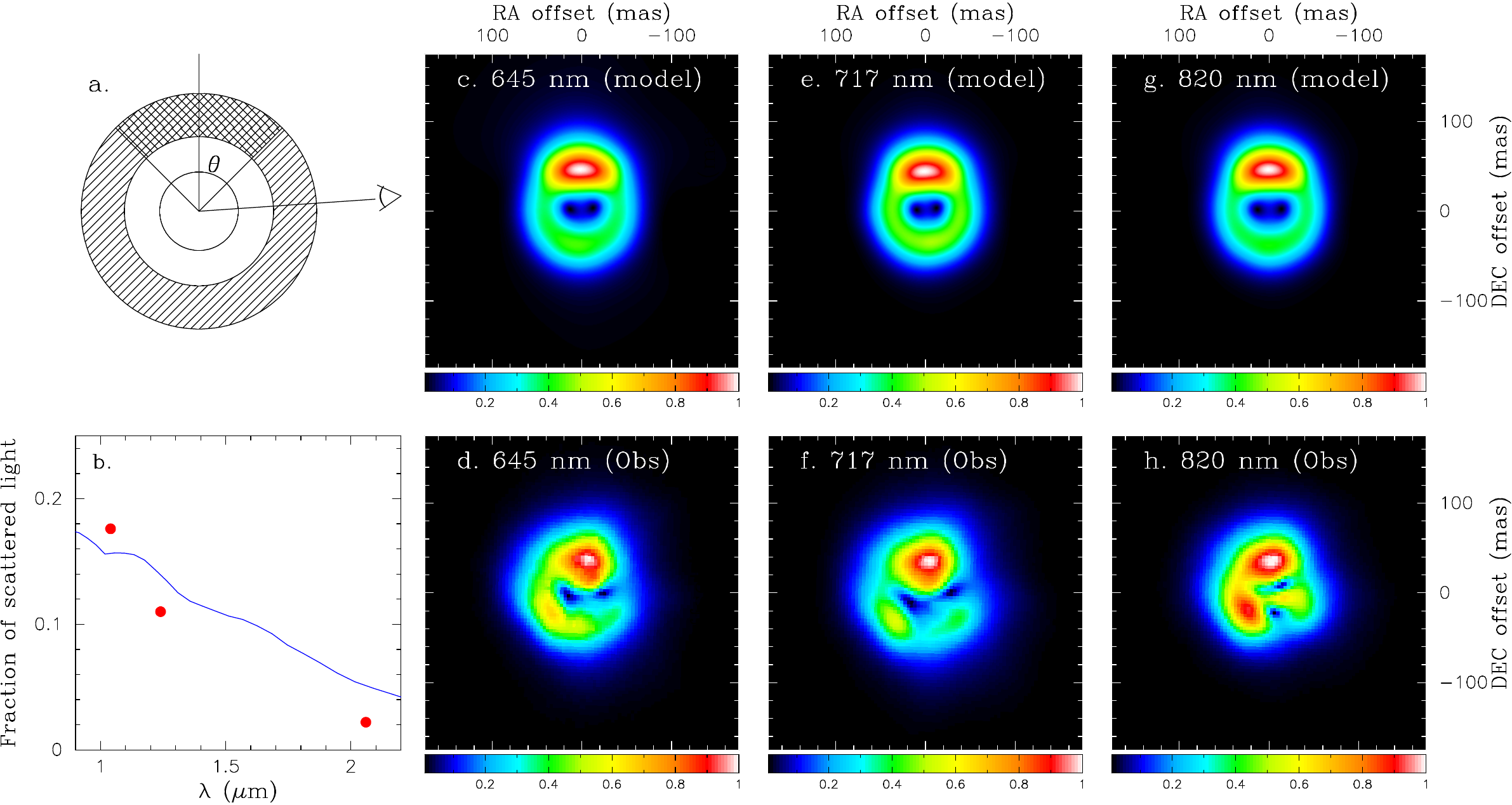}}}
\end{center}
\caption{
Dust clump model of \whya.  
{\bf a:} Schematic view of our model.  The hatched region is the spherical 
dust shell. The cross-hatched region 
represents a cone-shaped density enhancement, which is characterized 
by the half-opening angle $\theta$.  The viewing angle of the model 
is also shown.  
{\bf b:}  Fraction of scattered light derived by 
Norris et al. (\cite{norris12}) is plotted by the red dots (the errors are 
approximately the same as the size of the dots), 
while the model prediction is shown by the blue solid line. 
{\bf c--h:}  Model and observed polarized intensity maps at 645, 717, and 
820~nm.  North is up, east to the left.  
}
\label{clumpy_model_corundum}
\end{figure*}

However, as Fig.~\ref{whya_zimpol_1Davg_Halpha}a shows, 
the \mbox{1-D} intensity profile of our PSF reference star HD121653 
shows a halo that is more extended than that of \whya\ at intensity 
levels lower than 
1\% of the central peak (at angular distances greater than 120~mas), 
although HD121653 should appear as a point source. 
This means that the performance of AO was much worse for HD121653 than for 
\whya\ due to the worse seeing (see in Sect.~\ref{subsect_obs_zimpol}). 
If we convolve the model $I$ images with the PSF from HD121653, 
this extended halo in the PSF---despite its low intensity---makes the 
convolved $I$ images much more extended than it should be with the true, 
narrower PSF for \whya.  
When the convolved $I$ images are too extended, the predicted degree of 
polarization is significantly lower because of the division with $I$ in 
$\PL = \IP / I$, which makes a comparison  with the observed data impossible. 
However, we found out that the polarized intensity maps 
normalized with the peak value at each wavelength are not very sensitive to 
the extended halo of the observed PSF.  
Therefore, we used the 
normalized polarized intensity maps to constrain the properties of the 
inner dust envelope. 

\begin{figure*}
\resizebox{\hsize}{!}{\rotatebox{0}{\includegraphics{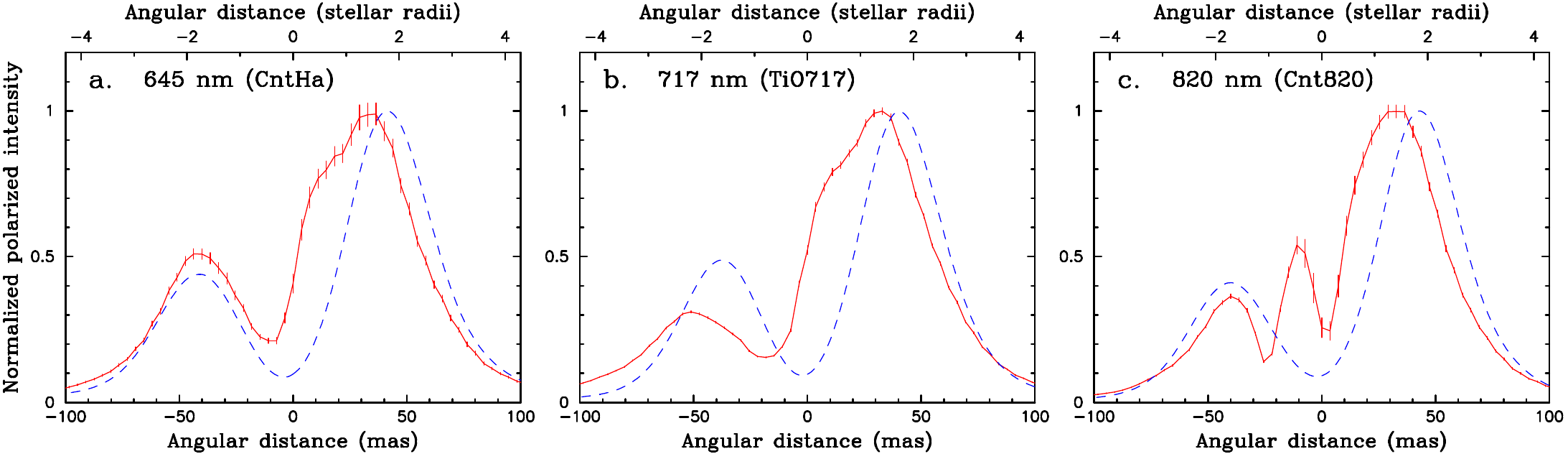}}}
\caption{
North-south 1-D cuts of the polarized intensity of \whya\ at 645~nm 
(panel {\bf a}), 717~nm (panel {\bf b}), and 820~nm (panel {\bf c}).  
In each panel, the red solid line represents the observed data, while 
the blue dashed line represents the model.  
}
\label{clumpy_model_corundum_1Dcut}
\end{figure*}

In our modeling, we computed the polarized intensity maps at the wavelengths 
observed with SPHERE-ZIMPOL and also the fraction of scattered light in 
the near-IR wavelengths studied by Norris et al. (\cite{norris12}). 
While our visible polarimetric imaging data allow us to constrain the 
properties of the innermost circumstellar environment, they cannot constrain 
the properties of grain species that give rise to the IR features.  
Therefore, we did not attempt to fit the SED.  
As mentioned in Sect.~\ref{sect_intro}, Khouri et al. (\cite{khouri15}) 
carried out a detailed modeling of the SED, including various dust features. 

The radiation of the central star was approximated with a blackbody of 
2500~K as in Khouri et al. (\cite{khouri15}).  
We adopted an angular diameter of 46.6~mas measured in the continuum with 
AMBER, which results in a radius of 383~\RSOL\ and a luminosity of 5130~\LSOL\ 
when combined with the distance of 78~pc from Knapp et al. (\cite{knapp03}).  

To explain the clumpy structure seen in the polarized intensity maps, 
we considered a dust shell model with a density enhancement, 
which is defined by a cone as depicted in Fig.~\ref{clumpy_model_corundum}a.  
The inner boundary of the dust shell was set to be the radius at which 
the dust temperature reaches a condensation temperature of 1500~K.  
We assumed the radial density distribution to be $\propto r^{-3}$ for the 
following reason.  
While the density distribution proportional to $r^{-2}$ corresponds 
to a stationary mass loss with a constant velocity, the density gradient 
in the innermost region of the envelope is expected to be steeper 
because the wind speed should increase at the base of the stellar wind.  
This is supported by the mid-infrared interferometric observations 
of Mira stars by Karovicova et al. (\cite{karovicova13}).  These data 
probe the inner region of the circumstellar envelope, and their modeling 
shows that the power-law index of the density distribution of \corundum\ is 
2.5--2.9. 

The free parameters in our clump model are the optical depth in the visible 
(550~nm) in the radial direction, the outer boundary radius of the shell, 
the half-opening angle of the cone of the density enhancement, and 
the ratio of the density in the cone and in the remaining region of the
shell.  
We considered three grain species, corundum (\corundum), 
forsterite (\forsterite), and enstatite (\enstatite), 
because they are thought to survive at high temperatures thanks to their 
low opacity in the visible and near-IR. 
The absorption and scattering cross sections, as well as the scattering 
matrix elements, were computed with the code of 
Bohren \& Huffman (\cite{bohren83}) for spherical grains, 
using the complex refractive index measured by Koike et al. (\cite{koike95}) 
for \corundum\ and the measurements of J\"ager et al. (\cite{jaeger03}) 
for \forsterite\ and \enstatite.  
We also computed models with the complex 
refractive index of \forsterite\ and \enstatite\ measured by Scott \& Duley 
(\cite{scott96}) to examine possible effects of different measurements.

Figure~\ref{clumpy_model_corundum} shows the \IP\ maps at 645, 717, 
and 820~nm predicted by the best-fit model with 0.5~\micron\ 
\corundum\ grains together with the observed data.  
This model is characterized by a density enhancement with a 
half-opening angle of 45\degr\ and a density ratio of 4 
within and outside the cone.  The inner and outer 
boundary radius is 1.9 and 3~\RSTAR, respectively, and the 550~nm optical 
depth is 0.1.   
As Fig.~\ref{clumpy_model_corundum}a shows, the viewing angle measured 
from the symmetry axis is 85\degr.  The density enhancement manifests 
itself as asymmetry in the \IP\ maps.  The ratio of the polarized 
intensity measured on the brightest clump in the north and the faintest 
clump in the SW in the SPHERE-ZIMPOL data is 2.7, 2.5, and 1.8 at 
645~nm, 717~nm, and 820~nm, respectively.  
The model predicts the ratio to be 2.3, 2.1, and 2.4 at 645, 717, and 820~nm, 
respectively, which agree with the observed ratios. 
Figure~\ref{clumpy_model_corundum_1Dcut} shows the \mbox{1-D} cuts of 
the observed and model polarized intensity in the north-south direction.  
The figure shows that our model can reproduce the observed ratio 
of the polarized intensity peaks reasonably well 
given the simplifications adopted in the model. 
However, the polarized intensity at the center of this model is much lower 
than the observed data.  We assumed a density enhancement in only one 
direction, while we detected three clumps possibly with different density 
enhancements.  Therefore, density distributions that are more complex than 
assumed here may reconcile the discrepancy in the polarized intensity near 
the center.

Figure~\ref{clumpy_model_corundum}b shows a comparison of the predicted 
fraction of scattered light at 1.04, 1.24, and 2.06~\micron\ and the observed 
values from Norris et al. (\cite{norris12}).  
The fraction of scattered light predicted by the model  agrees fairly well 
with the observed values, although the model predicts that the fraction 
should be higher than the observed value by a factor of 
$\sim$2 at 2.06~\micron. 
However, Norris et al. (\cite{norris12}) assumed a geometrically thin 
spherical shell to derive the fraction of scattered light.  Furthermore, 
the variability phase at the time of their observations was 0.2 
(post-maximum light), in contrast to the phase 0.9 of our SPHERE 
observations.  These factors may explain the discrepancy of a factor of 2 
at 2.06~\micron.

The parameters derived from our models with \forsterite\ or \enstatite\ 
are similar to those of the best-fit model with \corundum. 
The results obtained with the complex refractive indices of 
\forsterite\ and \enstatite\ measured by Scott \& Duley (\cite{scott96}) also 
agree with those obtained with the data of J\"ager et al. (\cite{jaeger03}). 
Our modeling with three grain species results in a 
550~nm optical depth of $0.1\pm0.02$ with grain sizes of 
0.4--0.5~\micron, and an inner and outer boundary radius of 1.9--2.0~\RSTAR\  
(defined by the assumed condensation temperature of 1500~K) 
and $3\pm0.5$~\RSTAR.  
The density enhancement is characterized by a half-opening angle of 
45\degr $\pm$15\degr\ and a density ratio of $4 \pm 1$.  
For the half-opening angle of 45\degr, viewing angles between 60\degr\ and 
100\degr\ (measured from the symmetry axis as shown in 
Fig.~\ref{clumpy_model_corundum}a) 
reproduce the observed ratios of the polarized intensity between the 
brightest and faintest clumps. 
As we explain in the next section, the models with a grain size smaller than 
$\sim$0.3~\micron\ cannot reproduce the observed polarized intensity maps. 
On the other hand, if the grain size is larger than $\sim$0.6~\micron, the 
predicted fractions of scattered light at 1.04, 1.24, and 2.06~\micron\ are 
much higher than the observed values from Norris et al. (\cite{norris12}).

\section{Discussion}
\label{sect_discuss}

The grain size of 0.4--0.5~\micron\ derived from our modeling 
is larger than the 0.3~\micron\ derived by Norris et al. (\cite{norris12}).  
If the grain radius of 0.3~\micron\ is adopted in our model, 
the 550~nm optical depth 
should be increased to 0.5 to explain the observed fraction of scattered 
light from 1.04 to 2.06~\micron.  Then the optical depth along the line 
of sight grazing the inner boundary of the dust shell becomes much larger 
than 1, even if the optical depth in the radial direction is still below 1.  
Multiple scattering for an optical depth larger than 1 
leads to very low polarization at the inner boundary, 
which is not observed in our SPHERE-ZIMPOL data.  
However, given the differences in the observational data, variability phase, 
and the assumptions in the models, the difference between the 
0.3~\micron\ derived by Norris et al. (\cite{norris12}) and the 
0.4--0.5~\micron\ from our modeling does not seem to be serious.  
The dust mass from our model is 
$5.8\times10^{-10}$~\MSOL\ with a bulk density of 4~g~cm$^{-3}$ adopted 
for \corundum. 
This is comparable to the $(1.09\pm0.02)\times10^{-9}$~\MSOL\ estimated by 
Norris et al. (\cite{norris12}), despite the difference in the grain size. 
The dust mass from our model also agrees well with $4.9\times10^{-10}$~\MSOL\ 
derived by Khouri et al. (\cite{khouri15}), who also assumed a grain size 
of 0.3~\micron.  
Therefore, our modeling based on the SPHERE-ZIMPOL polarimetric images 
confirms the predominance of large, transparent 
grains in \whya\ very close to the star, $\sim$2~\RSTAR, and reveals that 
the grain size is even larger than derived by Norris et al. (\cite{norris12}).

H\"ofner (\cite{hoefner08}) presents dynamical models with the mass loss 
driven by the scattering of stellar photons.  The parameters of these 
models (stellar mass, luminosity, effective temperature, and pulsation period) 
are comparable to those of \whya, if not perfectly optimized for \whya. 
The models predict that the grain size 
can reach 0.36--0.66~\micron\ at 2--3~\RSTAR.  
The grain size of 0.4--0.5~\micron\ as well as the inner radius of the dust 
shell derived from our modeling agrees with these predictions, 
lending support to the scenario that the scattering due to large, transparent 
grains can drive the mass loss in oxygen-rich AGB stars. 

In the dynamical models of H\"ofner (\cite{hoefner08}), however, 
the nucleation of dust grains from the gas phase is not considered. 
While the growth of \forsterite\ grains is followed in a time-dependent 
manner, the presence of seed nuclei is assumed, and their amount is treated 
as a free parameter.  
\mbox{Gobrecht} et al. (\cite{gobrecht16} and 
references therein) present comprehensive models for non-equilibrium 
chemical processes of gas and dust in the inner wind of AGB stars, 
incorporating the nucleation of dust grains from the gas phase, 
although dynamical aspects such as the radiation pressure on dust grains 
and the mass loss are not included. 
Their models predict that \corundum\ forms within 2~\RSTAR, 
and that the grain radius reaches 0.3~\micron\ in some cases. 
These model predictions are consistent with our results, given that the models 
of \mbox{Gobrecht} et al. (\cite{gobrecht16}) are optimized for the Mira star 
IK~Tau, which is more evolved than \whya.  
Their models predict that \forsterite\ forms beyond 3~\RSTAR\ and that 
\enstatite\ is far less abundant than \forsterite. Therefore, \corundum, 
rather than iron-poor silicates, may be the more plausible constituent of 
the clumpy dust clouds detected in our SPHERE-ZIMPOL observations.
Adjusting these models for \whya\ would allow us to 
test the non-equilibrium chemistry using the present data.

The observed polarized intensity maps suggest that a few large dust clumps 
nearly cover the entire sphere. 
In the case of the model shown in Fig.~\ref{clumpy_model_corundum}, 
the solid angle of the density enhancement is $4\pi \times 0.15$~str, 
which means that approximately seven clumps can cover the entire sphere.  
Freytag \& H\"ofner (\cite{freytag08}) present 3-D convective simulations 
for AGB stars with dust formation.  Their models show that the stellar 
surface is covered by a few, large convective cells and that the star 
pulsates with a typical time scale of one year.  
Dust forms behind the shock fronts associated with the pulsation and the 
convective cells.  
The dust formation region appears nearly spherical but with noticeable 
irregularities of the size of the convective cells.  
Although their 3-D models assume a carbon-rich chemistry unlike that of \whya, 
a similar phenomenon may be expected in oxygen-rich cases as well.  
The polarized intensity maps obtained with SPHERE-ZIMPOL show similar 
signatures---dust formation in an incomplete shell with clumpy structures.  
Therefore, our SPHERE-ZIMPOL observations lend support to the dust 
formation associated with the shocks induced by the pulsation and convection.

\section{Concluding remarks}
\label{sect_concl}

We have presented visible polarimetric imaging observations of the AGB 
star \whya\ with SPHERE-ZIMPOL and high spectral resolution 
($\lambda/\Delta \lambda = 12000$) interferometric 
observations with VLTI/AMBER in the individual CO first overtone lines near 
2.3~\micron.  
The polarized intensity maps obtained at five wavelengths between 645 and 
820~nm with spatial resolutions of 23--30~mas 
reveal three clumpy dust clouds close to the star, at $\sim$50~mas 
($\sim$2~\RSTAR).  
The continuum-subtracted \Ha\ image 
shows asymmetrically extended emission up to $\sim$160~mas, 
implying the propagation of shocks up to $\sim$7~\RSTAR.

The VLTI/AMBER observations have allowed us to spatially resolve the outer 
atmosphere of \whya.  
The fitting of the \mbox{AMBER} data in the continuum in the 2.3~\micron\ 
region with a power-law-type limb-darkened disk resulted in a 
stellar diameter of $46.6\pm 1.0$~mas with a limb-darkening parameter of 
$2.1\pm 0.2$.  
On the other hand, the AMBER data in the CO lines suggest that the atmosphere 
is extended to $\sim$70~mas ($\sim$3~\RSTAR) with Gaussian-like intensity 
distributions. 
Our high angular resolution observations with SPHERE-ZIMPOL and VLTI/AMBER 
reveals the coexistence of dust, molecular gas, and \Ha -emitting hot gas 
within 2--3~\RSTAR. 

Our Monte Carlo radiative transfer modeling suggests the presence of 
0.4--0.5~\micron\ grains of \corundum,\  \forsterite,\ or \enstatite\ 
in an optically thin ($\tau_{\rm 550nm} = 0.1\pm 0.02$) shell with 
an inner and outer radius of 1.9--2~\RSTAR\ and 
$3\pm 0.5$~\RSTAR, respectively.  
The grain size and the location of the dust formation 
is consistent with the hydrodynamical models with the mass loss driven by 
the scattering due to large grains.  
The clumpy structures detected in 
the SPHERE-ZIMPOL polarimetric images lend support to the 3-D simulations, 
in which dust forms behind the shock fronts associated with pulsation and 
large convective cells.  

Our SPHERE-ZIMPOL observations took place at pre-maximum light.  
Given the clear periodicity in the light curve of \whya, monitoring 
observations with SPHERE-ZIMPOL following the variability phase is 
extremely important in order to understand the role of pulsation 
in dust formation 
and mass loss.  Moreover, the velocity-resolved imaging with VLTI/AMBER 
taking advantage of its high spatial and high spectral resolution enables us 
to probe the gas dynamics in a model-independent manner.  
The velocity-resolved imaging in the CO lines and \HOH\ lines is ideal 
for detecting the initial acceleration of gas within $\sim$3~\RSTAR\ 
and, therefore, is indispensable for clarifying the driving 
mechanism of the mass loss.

\begin{acknowledgement}
We thank the ESO Paranal team for supporting our SPHERE and AMBER 
observations and Henning Avenhaus for helping us 
optimize the instrumental set-up of our SPHERE observations. 
We are also grateful to Mario van den \mbox{Ancker} and 
Julien Girard for providing us with the information about 
the orientation of the SPHERE-ZIMPOL detectors and the information about 
the data files with parameters 
relevant to the AO performance including the Strehl ratios.  
This research made use of the \mbox{SIMBAD} database, 
operated at the CDS, Strasbourg, France.  
We acknowledge with thanks the variable star observations from the AAVSO 
International Database contributed by observers worldwide and used in 
this research.
\end{acknowledgement}

\end{document}